\documentclass[aip, amsmath,amssymb,graphicx, preprint]{revtex4-1}

\usepackage{graphicx}
\usepackage{xspace}
\newcommand{\yso}[0]{Y$_2$SiO$_5$\xspace}
\newcommand{\eryso}[0]{{Er$^{3+}$:Y$_2$SiO$_5$}\xspace}
\newcommand{\eT}{\exp\left(-h \Delta_g/k_BT\right)}
\usepackage{braket}

\begin{document}


\title{Boltzman optical thermometry for cryogenics}



\author{Marek Zeman}
\affiliation{Univ. Grenoble Alpes, CNRS, Grenoble INP, Institut N\'eel, 38000 Grenoble, France}
\affiliation{Absolut System, 3 avenue Raymond Chanas, 38320 Eybens, France}

\author{Philippe Camus}
\affiliation{Univ. Grenoble Alpes, CNRS, Grenoble INP, Institut N\'eel, 38000 Grenoble, France}

\author{Thierry Chanelière}
\email[Corresponding author: ]{thierry.chaneliere@neel.cnrs.fr}
\affiliation{Univ. Grenoble Alpes, CNRS, Grenoble INP, Institut N\'eel, 38000 Grenoble, France}



\date{\today}

\begin{abstract}
We propose and implement an optical technique to access the local temperature of an erbium doped crystal by probing the electron spin population under magnetic field. We reliably extract the sample temperature in the range 2\,-\,7\,K. We additionally discuss the suitability of our method as a primary standard for cryogenic thermometry. By adding an auxiliary heating laser, we are able to measure the interface conductance between the dielectric crystal and the cold plate of the cryostat by exploring different cooling configurations.
\end{abstract}

\pacs{}

\maketitle 

\section{Introduction}

Starting at room temperature, optical thermometry using luminescent local probes continues to gain popularity, now significantly amplified by its prospects in biology and healthcare. It relies on a wide range of materials—organic molecules, semiconductors, metallic or insulating nanoparticles with emitting impurities (lanthanides, colored diamond centers, transition ions) \cite{Bradac2020} — whose emissions are compared at least at two wavelengths or by measuring the radiative lifetime, seeking for the highest thermal sensitivity of the phenomenon \cite{Wang2013}.

Lanthanides hold a special place as they easily integrate into inert inorganic matrices compatible with biological tissues \cite{Brites2018}. They also withstand high temperatures (up to $2000$\,K), both to characterize an extreme environment where inserting a thermometer is impossible or to induce strong, localized heating of tissues \cite{ZHONG201834, Abbas2022, Er_2000K, D3TC02043F}.

The recent emergence of new applications has expanded the range of parameters to explore \cite{Harrington2023}. This includes cryogenic temperatures for monitoring superconducting magnets, space applications, and cryo-electron microscopy of biological samples for example, especially as commercial cryomachines are now widely available, robust and efficient.

The race to ultra-low temperatures has strongly stimulated the need for precision thermometry, covering the range from the ambient to the millikelvin temperatures, spanning several orders of magnitude \cite{Liu2015,Mykhaylyk:vv5165,Back2020}. The question of establishing a primary or secondary standard is recurring regardless of the application \cite{Pobell2007, Dedyulin2022}. We will also discuss this point in our case.

We propose using an erbium-doped dielectric crystal to explore the restricted range around 4\,K, where many cryogenic devices operate. Our goal is not to investigate a wide range but to provide an optical probe of the crystal's temperature. Like many cooled insulator substrates (like sapphire or silicon oxide for example), the question of the exact sample temperature compared to the cold plate where the reference thermometer is attached inevitably arises. It is therefore interesting to have a local probe with a contactless measurement.

The success of lanthanides in thermometry ultimately comes from the abundance of levels whose energy differences are comparable to the temperature $T$ (in units of $k_B T$, where $k_B$ is the Boltzmann constant). This intrinsic property of electronic levels makes spectroscopic measurements sensitive to temperature, which affects in turn the level population and modifies the emission or the absorption spectra. We propose to extend this logic to the 4\,K temperature range using electronic spin levels whose splitting under a magnetic field (divided by $k_B$) can be adjusted to a few kelvins. We then perform an optical measurement of the spin polarisation, i.e., a ratiometric measurement in the sense of optical thermometry.

In this regard, our method has a kinship with magnetic thermometry, where nuclear polarisation is measured by magnetic resonance. It is also comparable to the 
tremendous work on diamond NV centers \cite{Kucsko2013}, sometimes called quantum thermometry because of the remarkable coherence properties of a unique center in nanodiamond \cite{doi:10.1146/annurev-physchem-040513-103659,https://doi.org/10.1002/cnma.201700257, Fujiwara_2021}.  In a similar manner, we perform the optical reading of electronic spin resonance levels, even though we are not seeking spatial resolution as compared to NVs.

We propose to use a earth-rare ion doped crystal. More specifically, using erbium as a dopant. Erbium has an electron spin that can be measured optically thanks to the narrow transition linewidth. The electronic spin splitting under a moderate magnetic field is comparable to our temperature range of interest.
Erbium has a particularly large electron spin due to the multiplicity of electrons in the 4f shell. Although its g-factor can vary from 2 to 15 typically, due to the strong anisotropy of the dipole moment, a median value of 7 can be used, which corresponds to a gyromagnetic splitting of 100\,GHz/T. By applying a moderate local field of 0.5\,T for example, the splitting will be 2.3\,K (in units of $k_B T$). So a moderate thermal variation in this range will therefore induce a significant Boltzman population change that can be measured using optical methods. This range of magnetic fields is accessible by the use of permanent magnets, enabling compact assembly that can be transported from one platform to the another to meet the demand.

We begin by describing the optical method for measuring the electron spin of erbium in section \ref{sec:methods}. High-resolution spectroscopy is used to measure the polarisation of the spin in the ground state and to deduce the local temperature by applying Boltzman's law. The sensitivity of the method will be analysed in section \ref{sec:sensibility}. As we shall see, it is not possible to make an absolute temperature measurement, despite the universality of Boltzman's law. We will discuss the reasons for this in section \ref{sec:primary}. The method does, however, define a secondary standard which, after calibration, can be used to measure in-situ the temperature of the dielectric crystal. This will be applied to measure the heat conduction from the crystal heated by an external laser beam to the cold plate in two cryostats using different types of thermal joints (silver lacquer, varnish, grease) as described in section \ref{sec:conduction}.

\section{Methods}\label{sec:methods}
\subsection{Boltzman optical thermometry of the electron spin}
Our method consists of measuring the spin polarisation (population) whose splitting under magnetic field corresponds to the temperature range of interest, around 4K in our case. As outlined in the introduction by an order-of-magnitude calculation, earth-rare ions are well suited for this purpose. Not only are the Zeeman splittings compatible with our objective, but above all the spin transitions are narrow enough to be resolved by optical spectroscopy \cite{liu2005spectroscopic}. Without prejudging the generalisation of our method to other dopants, we have chosen to use erbium, which has two additional advantages in the lanthanide series. Firstly, it has a high gyromagnetic factor, which reduces the required magnetic field. Secondly, its transition is in the range of fibre telecommunications where DFB lasers are spectrally narrow, widely available, small and cheap.

\begin{figure}[ht!]
\centering
\includegraphics[width=\columnwidth]{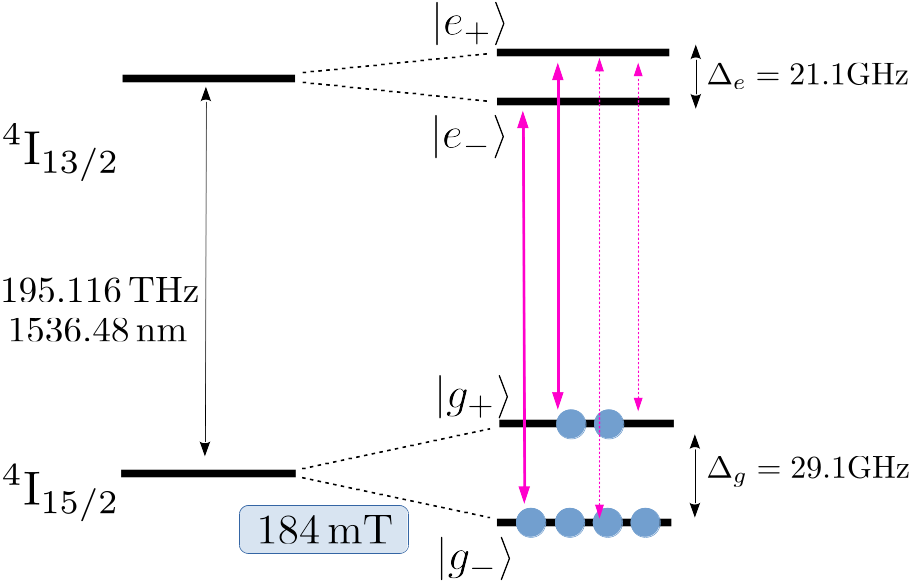}
\caption{The ground and optically excited states split under magnetic field (184\,mT for example) into the two components of a Kramers doublet noted $\ket{g_-}$,$\ket{g_+}$ and $\ket{e_-}$,$\ket{e_+}$ respectively. The level scheme is restricted to the lowest lying crystal field levels.  The splittings are respectively noted $\Delta_g$ and $\Delta_e$ with the corresponding values at 184\,mT. The $\ket{g_-} \rightarrow \ket{e_-}$ and $\ket{g_+} \rightarrow \ket{e_+}$ are called the direct transitions (solid magenta arrows) and the $\ket{g_-} \rightarrow \ket{e_+}$ and $\ket{g_+} \rightarrow \ket{e_+}$ are the crossed transitions  (dashed magenta arrows).}
\label{fig:levels}
\end{figure}

At low temperatures, the structure of erbium in \yso can be represented schematically in Fig.\ref{fig:levels} when restricted to the transition of interest. We choose \eryso that has the advantage to be well characterized in the literature \cite{Bottger2006}. The $^4\mathrm{I}_{15/2} \rightarrow ^4\mathrm{I}_{13/2}$ transition is addressed here, and more specifically the two lowest energy levels of the crystal field multiplets. The fundamental and excited levels are strongly anisotropic electron spin doublets when the magnetic field is oriented in the crystal frame. The structure of interest here can be summarised as four levels, two split doublets, linked by the optical transition at 1536.48\,nm. The g-factors in the ground and excited states for our orientation are $g_g=11.2$ and $g_e=8.2$ respectively \footnote{We address the so-called site 1
in Böttger’s work \cite{Bottger2006} and use the same orientation as in \cite{Dajczgewand:14}. The laser polarisation is aligned along the so-called D$_2$ extinction axis to maximise the absorption}.To put things in perspective, with a typical value for the magnetic field we use, $B=184$\,mT, the splittings are respectively $\Delta_g=g_g \mu_B B=29.1$\,GHz and $\Delta_e=g_e \mu_B B=21.1$\,GHz ($\mu_B$ is the Bohr magneton expressed in Hz/T).

\begin{figure}[ht!]
\centering
\includegraphics[width=\columnwidth]{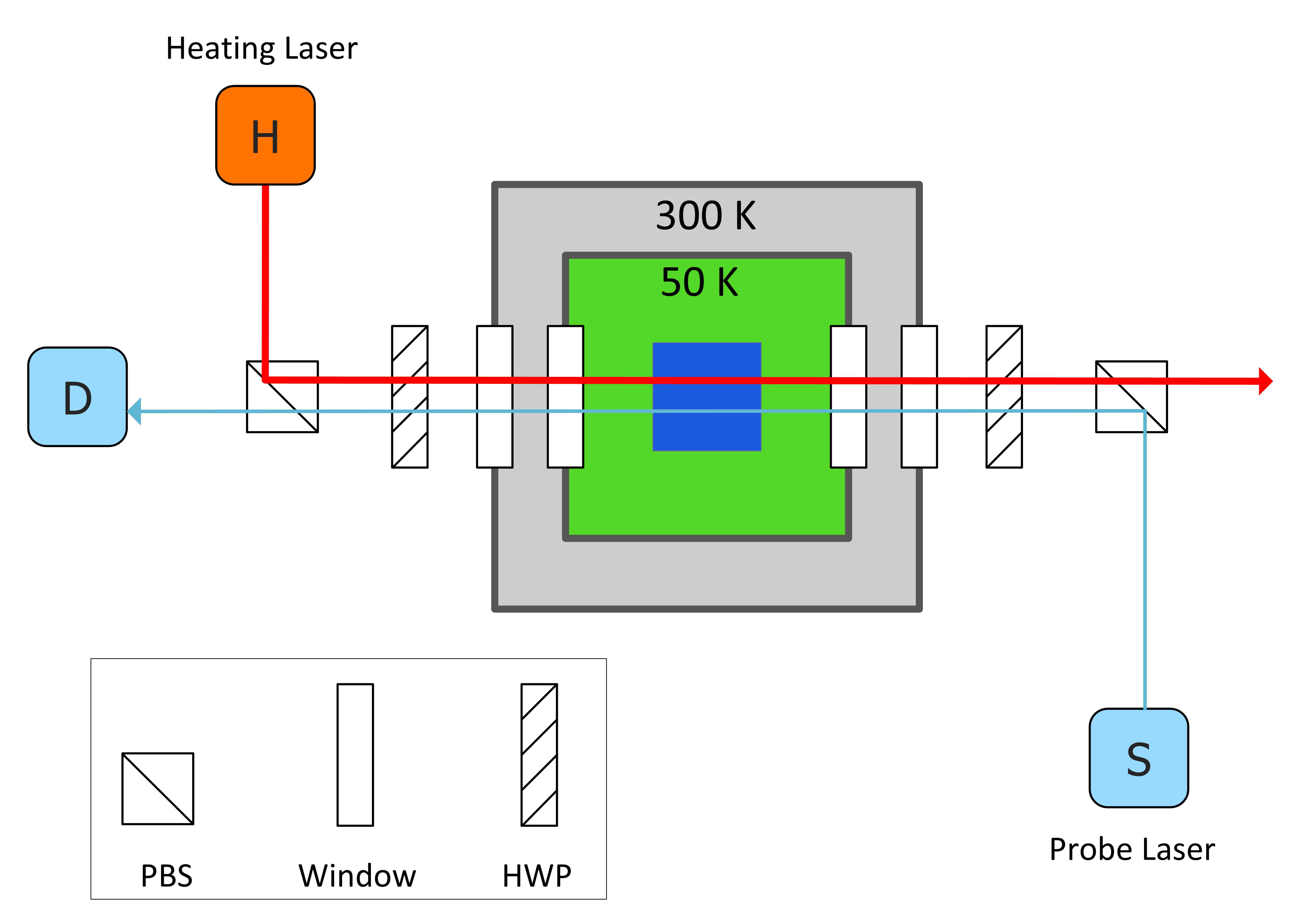}
\caption{The optical setup is composed of two lasers. We record the absorption spectra as plotted in Fig.\ref{fig:spectra} from the frequency-swept probe laser to measure the ratiometric temperature. We also add a independent off-resonant heating laser that will be used in section \ref{sec:conduction} to evaluate the sample heat load.}
\label{fig:mount}
\end{figure}

The Boltzman law applied to the ground state doublet $\ket{g_-}$ and $\ket{g_+}$ imposes the population to be  $\displaystyle \frac{1}{1+\eT}$ and $ \displaystyle \frac{\eT}{1+\eT}$ if the total population is normalised to one. The population of the states scales the absorption of the different transitions that can be measured by optical spectroscopy. The optical setup is depicted in Fig.\ref{fig:mount}

\subsection{Measurement method and temperature calibration}\label{calibration}

\begin{figure}[ht!]
\centering
\includegraphics[width=.5\columnwidth]{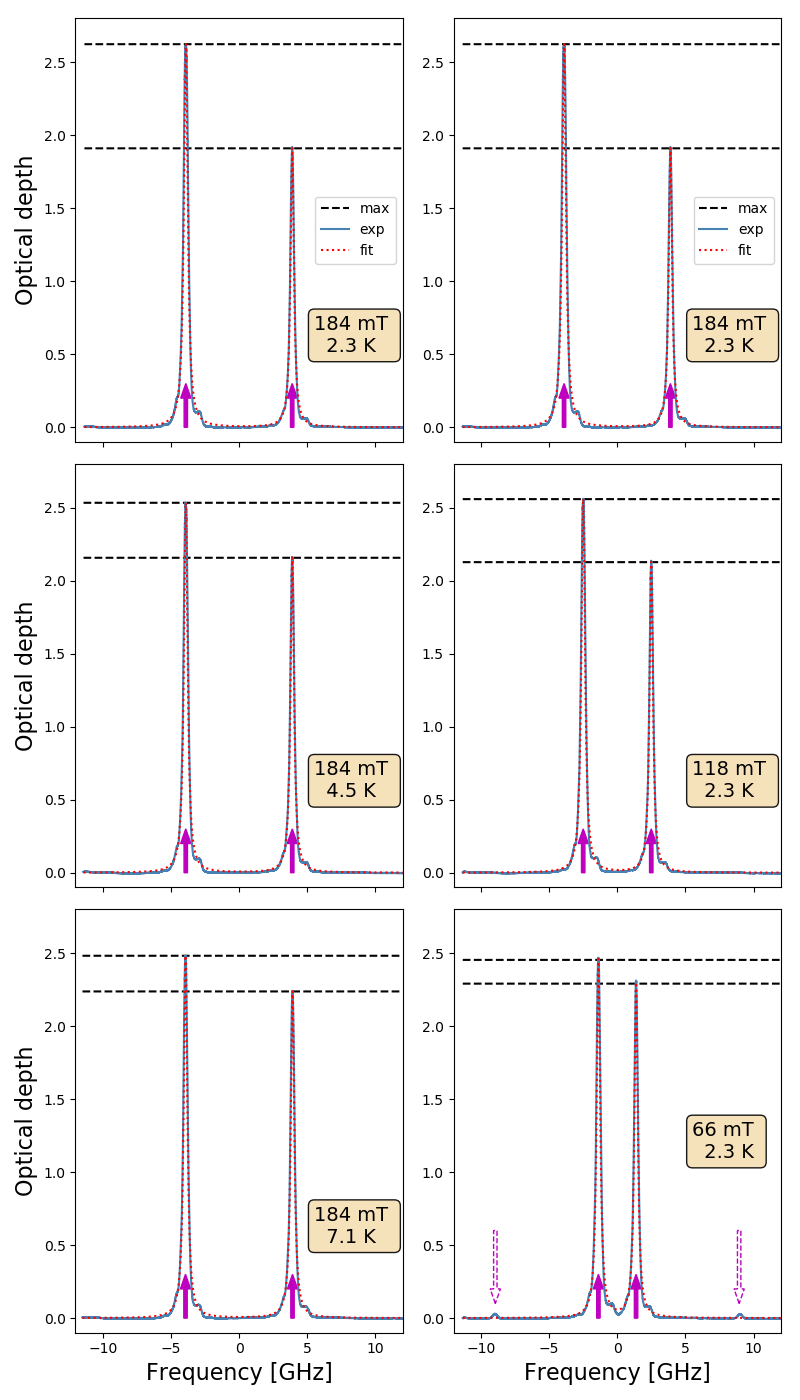} 
\caption{Absorption spectra illustrating the Boltzman law. The left column corresponds to an increasing temperature (from top to bottom at fixed magnetic field), and the right column when decreasing magnetic field (from top to bottom at fixed temperature). The absorption is expressed as the natural log of the transmission (optical depth). The black dashed lines mark the maxima of the peaks that are captured by a Lorentzian fit (red dotted line). Each peak is systematically flanked with a pedestal visible on its right side. It is explained by the natural abundance (23\%) of the $^{167}$Er isotope. The latter has a zero-field hyperfine structure that cannot be simply described by a split doublet as in Fig.\ref{fig:levels}. The Lorenztian curve chosen to fit the peaks thus represents an average shape of the absorption line.}
\label{fig:spectra} 
\end{figure}

We record the absorption spectrum through the crystal by sweeping a narrow band laser (Velocity\textsuperscript{\textregistered} External Cavity Diode Laser) around the transition of interest. The spectrum analysis actually reveals the population balance and in turn the temperature as illustrated in Fig. \ref{fig:spectra}. The main peaks corresponds to the direct transitions depicted in Fig.\ref{fig:levels} and are marked with magenta arrows in Fig. \ref{fig:spectra}. They are positioned at $\pm (\Delta_g-\Delta_e)/2$ originating from the ground substates $\ket{g_\pm}$. The absorption of the transition from the $\ket{g_-}$ is the strongest because of the Boltzman law. The crossed transitions are not visible because they are weaker and positioned at $\pm (\Delta_g+\Delta_e)/2$. So they are generally out of our laser frequency scan except for the lowest magnetic field, 66\,mT, where the crossed transitions are marked with dashed magenta arrows in Fig. \ref{fig:spectra}.

We focus on the direct transitions whose absorption (optical depth) will be used to evaluate the temperature. To do so, we fit a Lorentzian profile to each peak (red dotted line in Fig. \ref{fig:spectra}). This procedure convincingly capture the maxima of the absorption sprectrum (marked by the black dashed lines in Fig. \ref{fig:spectra}) that we call $A_-$ and  $A_+$ for $\ket{g_-} \rightarrow \ket{e_-}$ and $\ket{g_+} \rightarrow \ket{e_+}$ respectively. If  $A_-$ and  $A_+$ are indeed proportional to the populations in $\ket{g_-}$ and $\ket{g_+}$, this assumption will be discussed specifically in \ref{sec:primary}, applying the Boltzman law allows us to define the so-called ratiometric temperature as 

\begin{equation}\label{eq:T_ratio}
T_\mathrm{ratio}= \frac{h \Delta_g}{k_B} \frac{1}{\log (A_-/A_+)}
\end{equation}

\begin{figure}[ht!]
\centering
\includegraphics[width=.8\columnwidth]{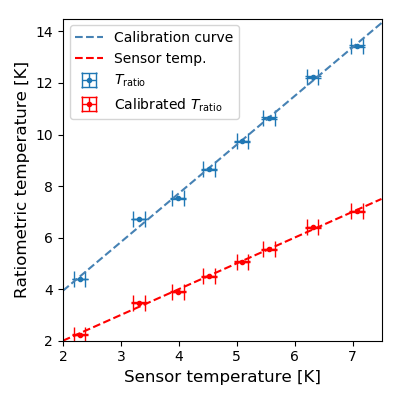} 
\caption{Ratiometric temperature $T_\mathrm{ratio}$ (blue markers) as a function of the cryostat sensor temperature. The latter is reported as a red dashed line for comparison. The difference cannot be explained by a poor thermalisation of the sample as discussed in the text. We use a linear fit to data as a calibration curve (blue dashed line) to relate the measured $T_\mathrm{ratio}$ and the cryostat temperature (red dashed line). The red markers represent the ratiometric temperature after calibration using the linear fit. The uncertainties (vertical error bars) will be discussed specifically in \ref{sec:sensibility}. The horizontal error bars are due to the uncalibrated sensor with a $\pm0.1$\,K accuracy.}
\label{fig:calibration}
\end{figure}

Having defined this temperature, we'd like to compare it with the temperature measured by an independent sensor (Lakeshore DT-670) in the cryostat. The latter, a Janis SVT wet cryostat, has a variable-temperature insert (VTI) whose temperature we vary between 2.3\,K (minimum cryostat temperature) and 7\,K, corresponding to our range of interest. The magnetic field is provided by a superconducting solenoid in the insert. The field magnitude is variable, which is useful for this calibration and analysis stage.
We limit ourselves in practice for the whole study to this range 2\,-\,7\,K. On one side of this range, because it corresponds to the minimum temperature of the cryostat used to evaluate our method. This is also the case for the test cryostat that we included in section \ref{sec:conduction}. Implementations at lower temperatures would be interesting, but would raise fundamental questions about the spin thermalisation time, as will be discussed briefly in conclusion \ref{ccl}. On the other side of the temperature range, beyond 7\,K, we lose sensitivity as we will discuss specifically in section \ref{sec:sensibility}. There is also a technical reason for the latter limit. We use a niobium–titanium superconducting magnet with a critical temperature of about 10\,K. Operating below 7\,K avoids the quenching of the coil.
We now compare the ratiometric temperature $T_\mathrm{ratio}$ and the sensor temperature in Fig. \ref{fig:calibration}.

The first thing to notice is that the ratiometric temperature differs significantly from the temperature in the cryostat insert. This is not a problem of sample thermalisation. To rule out this hypothesis, it should be noted that the crystal is efficiently cooled by the gas in the insert. Furthermore, we have verified that the temperature difference does not depend on the laser intensity (as a possible source of heating) or on the waiting time during the experimental run. We also expect the spin dynamics to be fast at this temperature and dopant concentration \cite{PhysRevB.100.165107}. The difference in temperature plotted in Fig. \ref{fig:calibration} is actually intrinsic to the method. The ratiometric temperature is actually not an absolute temperature, in other words, it is not a primary standard in the sense of metrology as we will discuss in detail in \ref{sec:primary}.
We will therefore use $T_\mathrm{ratio}$ as a secondary standard which must be calibrated first. Despite the temperature difference in the cryostat, there is a linear dependence of $T_\mathrm{ratio}$  in Fig. \ref{fig:calibration}, which we will use as a calibration curve (blue dashed line) for the rest of the work, namely the in-situ measurement of the crystal temperature as we will see in section \ref{sec:conduction}. Before doing so, we wish to discuss the limitations of our method in terms of uncertainties in \ref{sec:sensibility} and explain the difficulty of obtaining a primary standard from spectroscopic measurements in \ref{sec:primary}.

\section{Uncertainties and sensibility of the method}\label{sec:sensibility}

In order to evaluate the measurement uncertainty that represents the limit of our sensor accuracy, the impact of noise on the temperature prediction must be quantified.

The signal-to-noise ratio of the transmission curve is of the order of 10$^3$ as acquired by the oscilloscope. The fitting procedure to estimate $A_-$ and  $A_+$ further reduces the uncertainty by an order of magnitude. The error is canonically extracted by the fitting algorithm that provides the covariance matrix. We typically find an uncertainty of 10$^{-4}$, not very visible on the figure \ref{fig:calibration} and in any case negligible with respect to the precision of the  sensor temperature (10$^{-2}$) and its accuracy (10$^{-1}$ for a non-calibrated sensor). Regardless of the detail of the fitting algorithm, the final uncertainty can be estimated as the noise on the sampling measurement ($\sim$ 10$^{-3}$) divided the square root of the number of samples contained in the width of each peak $A_-$ and  $A_+$ ($\sim \sqrt{1000}$ in our case) which acts as an averaging interval for the fit.

This analysis makes it possible to question the sensitivity of the sensor more generally than under the experimental conditions chosen for our demonstration. The aim here is to widen its scope and explore different working ranges.
The sensitivity of the measurement will depend strongly first on the magnetic field chosen for the temperature range to be explored. In the introduction, we have intuitively proposed to choose the magnetic field producing a level splitting comparable to our temperature of interest. This can be quantified as follows.

The temperature estimation comes literally from the comparison of amplitudes $A_-$ and  $A_+$. At a given temperature, if the magnetic field is too weak, the peaks are comparable and difficult to differentiate in the measurement noise limit. Conversely, in strong fields, peak $A_+$ will be totally overwritten and even difficult to measure in the dynamic range of the acquisition system. This compromise can be quantitatively translated by assuming that the standard deviation of the noise on the amplitude fitting $\sigma_A$ is the same for $A_-$ and  $A_+$, but the signal-to-noise ratio $\displaystyle \mathrm{SNR_A}= \frac{A_-}{\sigma_A}$ applies to the highest peak $A_-$ only as a reference for the measurement, and in turn the signal-to-noise ratio of the smallest peak $A_+$ is $\displaystyle \mathrm{SNR_A} \times \frac{A_+}{A_-}$ affected by the ratio $\displaystyle \frac{A_+}{A_-}$.

The uncertainty on the fit $\sigma_A$ propagates to the ratiometric temperature defined by Eq.\eqref{eq:T_ratio} leading to the error $\sigma_T$ on the temperature given by

\begin{eqnarray}
\frac{\sigma_T}{T} &&= \frac{1}{\mathrm{SNR_T}} = \frac{k_BT}{h \Delta_g}\left[1+\exp \left(\frac{h \Delta_g}{k_BT}\right) \right]\times \frac{\sigma_A}{A_-}\nonumber\\
&& = S_N \left(\frac{\Delta_g}{T}\right)\times\frac{1}{\mathrm{SNR_A}} \label{eq:sensibility}
\end{eqnarray}
where the noise ratio $\displaystyle \frac{\sigma_T}{T} = \frac{1}{\mathrm{SNR_T}}$ (inverse of the signal-to-noise ratio) is linked to $ \mathrm{SNR_A}$ by what we define as the noise sensibility $S_N$

\begin{equation}\label{eq:sensibility_S}
S_N \left(\frac{\Delta_g}{T}\right) =  \frac{k_BT}{h \Delta_g}\left[1+\exp \left(\frac{h \Delta_g}{k_BT}\right)\right]
\end{equation}

\begin{figure}[ht!]
\centering
\includegraphics[width=\columnwidth]{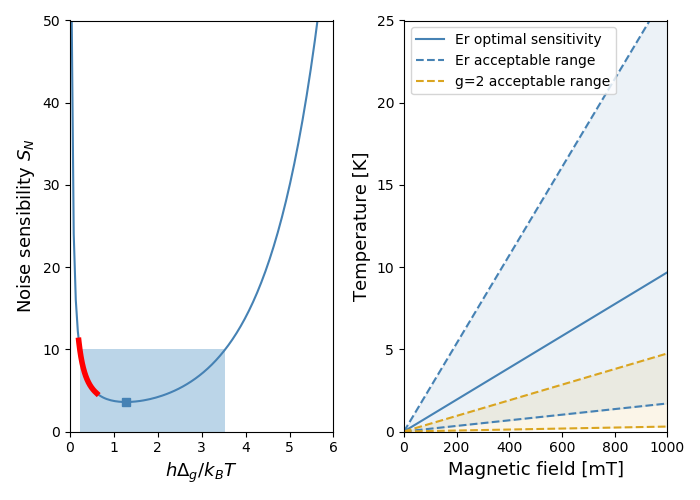} 
\caption{Left: Noise sensibility $S_N$ defined by uncertainty propagation from the raw measurement in \eqref{eq:sensibility}. We define the optimal conditions (blue square marker) and an acceptable range (blue shaded area) in the text. The thick red line represents the range of temperature that we have explore in Fig.\ref{fig:calibration}. Right: Illustration of the acceptable parameter range for which the noise sensibility stays below 10 for the erbium spin ($g_g=11.2$, blue shaded area) and a generic electron spin ($g=2$, orange shaded area). See the text for details.}
\label{fig:sensibility}
\end{figure}

This latter quantifies the noise amplification from the raw measurement of the amplitudes $A_-$ and  $A_+$ to the estimated ratiometric temperature. This factor varies by orders of magnitude depending on the choice of the splitting $\Delta_g$ relative to the temperature of interest $T$ as illustrated in Fig. \ref{fig:sensibility} (left).

The noise sensibility $S_N$ strongly varies as a function of $\Delta_g/T$. It reaches a minimum of $\simeq 3.59$ (blue square marker in Fig. \ref{fig:sensibility}, left) for $\displaystyle \frac{h \Delta_g}{k_BT} \simeq 1.28$. This allows us to quantify our intuition: the splitting (in units of $k_B T$) must be close to the temperature of interest. At this optimum, the noise measurement is only amplified by a factor of $3.59$ affecting the measured temperature. We propose to define an acceptable parameter range for which the noise sensibility stays below 10 (blue shaded region in Fig. \ref{fig:sensibility}, left) this corresponds to the following interval for $\displaystyle \frac{h \Delta_g}{k_BT} \simeq [0.23,3.54]$. This doesn't mean that the method cannot be used outside of this range, this simply means that the noise sensibility will be higher. This latter could be compensated by technical improvements to reduce the initial measurement noise and in turn to maintain the precision on the temperature.

In Fig. \ref{fig:sensibility} (left), we have also represented the range of temperature that we have explore in Fig.\ref{fig:calibration} (thick red line), namely 2.3K - 7K with $B=184$\,mT. This appears as a red thick segment. We are more or less in the acceptable parameter range, but below the optimum. Indeed, if we had wanted to obtain the best sensitivity at 4.2K for example, the necessary magnetic field would have been 554\,mT, which was not directly possible in our case (limited by our superconducting solenoid).

To be more specific, we finally propose to compare the optimal magnetic field for a given temperature of interest in Fig. \ref{fig:sensibility} (right). The optimal splitting, defined by the relation $\displaystyle \frac{h \Delta_g}{k_BT} \simeq 1.28$, will depend on the g-factor of the ion under consideration. Thus, we derive the temperature range that corresponds to the previously defined acceptable range of parameters as a function of the magnetic field, for erbium with $g_g=11.2$ in blue and for a generic electron spin with $g=2$ in orange. It is shown that erbium allows to explore the cryogenic temperature range up to about 30K with moderate magnetic field $<$1000mT (with permanent magnets for example). Using instead a generic electron spin with $g=2$ significantly reduces the accessible temperature range.

After discussing the precision of the method, we wish to discuss its accuracy, in the sense that the ratiometric temperature is not an absolute temperature.

\section{Suitability as a primary standard for thermometry}\label{sec:primary}

As discussed in \ref{calibration}, it is not possible to use the ratiometric temperature as a primary standard, despite the universal nature of the Boltzman's law. We have ruled out the question of sample thermalisation for practical reasons because of the efficient cooling by the gas. Furthermore, the spins whose population we are measuring have their own dynamics driven by phonons. However, the dynamics of the spins for these dopant concentrations and at these temperatures is relatively fast compared with the time of the experiment \cite{PhysRevB.100.165107}. So this is not an intrinsic limitation for the moment.
To explain the discrepancy between $T_\mathrm{ratio}$ and the cryostat temperature, it is useful to go back over the spectroscopy experiment in detail.
The absorption coefficient depends not only on the populations that allow us to extract the temperature, but also on the strengths of the transitions (in the sense of the transition dipole) as a multiplicative factor. We could have been convinced that the two direct transitions within two Kramers doublets have the same strength, given the similarity of the wave functions. The latter remain relatively pure as much as the level splittings under magnetic field remain small compared with the typical energy between crystal field levels. However, this simplifying assumption is not verified in practice, as the following experiment shows.

\begin{figure}[ht!]
\centering
\includegraphics[width=\columnwidth]{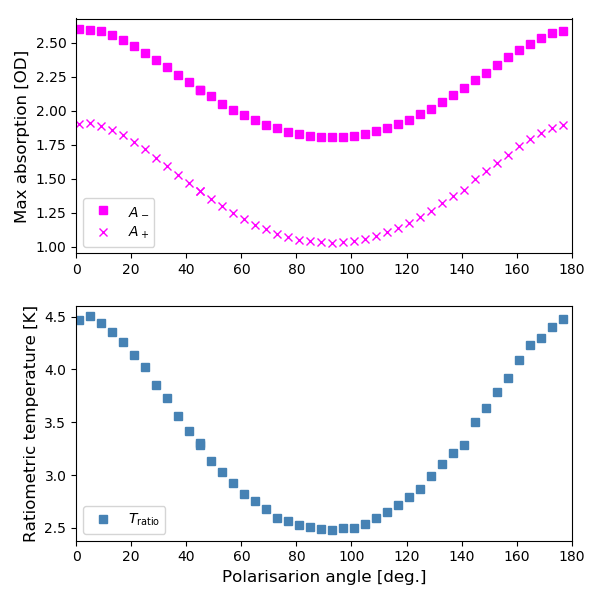} 
\caption{Top: Maxima of the absorption peaks $A_-$ and  $A_+$ as pointed out in Fig.\ref{fig:spectra} when we vary the incoming polarisation for a fixed cryostat temperature of 2.3\,K. We rotation the polarisation in the (D$_1$,D$_2$) plane where 0$^\circ$ and 90$^\circ$ corresponds to D$_2$ and D$_1$ respectively. Bottom: Corresponding ratiometric temperature $T_\mathrm{ratio}$ Eq.\eqref{eq:T_ratio}.}
\label{fig:hwp}
\end{figure}

To investigate the strengths of the direct transitions, for a given population balance (given temperature), we simply rotate the incoming polarisation and record again the absorption spectra and derive in turn the ratiometric temperature as shown in Fig.\ref{fig:hwp}.

We should not be surprised by the variation of the global absorption when the polarisation is varied. This anisotropic character is due to the monoclinic nature of the \yso
\cite{PETIT2020100062,liu2005spectroscopic}. What is more surprising is that the ratio of $A_-$ and  $A_+$ varies, as demonstrated by the variation of $T_\mathrm{ratio}$ (Fig.\ref{fig:hwp}, bottom). This fact actually shows that the strengths of the transition also depends on the polarisation. As a consequence, the ratio $A_- / A_+$ is not rigorously equal to the population ratio, but also contains a contribution from the independent transition dipoles. This deserves to be discussed.

It should be first noted that the variation in the absorption ratio in Fig.\ref{fig:hwp} only corresponds to a 10\,\% variation in the relative oscillator strengths. Our method, designed to be sensitive to population variations, is also sensitive to fluctuations in absorption from other sources.

Additionally, the evaluation of oscillator strength is an old question \cite{hufner_optical, liu2005spectroscopic}, recently reconsidered \cite{PETIT2020100062, PhysRevB.104.155121}, for which there is no general model, but only symmetry considerations on the nature, magnetic (MD) or forced electric dipole (ED), and the orientation of the dipoles. \eryso is a doubly complicated case in that sense. On the one hand, the symmetry of \yso is low. The orientation of the dipoles therefore has no obvious link with the crystal axis and must be represented by a tensor. On the other hand, the erbium transition is also magnetic dipole allowed ($\Delta J=-1$) and will in practice be a mixture of two MD and ED contributions \cite{Moncorge, Bottger_Spectroscopy}. This is a key point that should be sufficient to explain the difference between direct transitions. They connect the two sublevels of the Kramers doublets. In each doublet the wave functions are related by the time reversal operator \cite{hufner_optical}. MD and ED do not have the same parity (odd and pair, respectively) by time reversal \cite{AFZELIUS20101566}. As a result, the total sum of MD and ED contributions can be, for example, constructive for one of the direct transitions, but destructive for the other when applying time reversal. The two direct transitions can then have different strengths, thus explaining their different behaviour when rotating the polarization. A similar study on another transitions or ions where the MD transition is forbidden would allow an experimental verification.

Coming back to thermometry considerations, whatever is the exact origin of the previously discussed oscillator strength difference, we could argue that when the polarisation is at 90$^\circ$, corresponding the D$_1$ extinction axis, $T_\mathrm{ratio}\simeq 2.5$\,K is closer to the cryostat temperature of 2.3\,K. That is a fact, but a discrepancy persists, at the price of a lower absorption contrast because the absorption is weaker.
At the end, $T_\mathrm{ratio}$ can only be used as an indicator of the temperature that should be calibrated at given experimental conditions, namely the polarisation and the strength of the magnetic field.

Even if our method is limited in terms of obtaining a primary standard, it retains a strong interest that makes its originality.

\section{Evaluation of heat conduction by optical thermometry}\label{sec:conduction}

Optical thermometry allows a local measurement of the sample temperature. While cryogenic thermometers ensure excellent measurement accuracy on the sample holder, the exact temperature of the sample remains an open question, especially as the nature of the materials is very different, mostly metals or dielectrics (insulators).

As a proof of principle, we propose to evaluate the thermal conductance between the sample, which can be considered as a typical crystalline dielectric, and the sample holder or the surrounding exchange gas if present. This is an opportunity for us to test in a dry closed-cycle cryostat developed by Absolut System for optical applications (ACE-CUBE), whose cold plate under vacuum reaches 2.0\,K thanks to an additional Joule–Thomson cooling stage.

We therefore installed a second laser dedicated to heating the sample directly. This laser is also in the telecom range, but detuned by a few nm to avoid disturbing the spectroscopic probe that  $T_\mathrm{ratio}$ is measuring (see Fig.\ref{fig:mount}). It delivers up to a hundred mW. By varying its power and measuring the ratiometric temperature, we can characterise the thermal load on the crystal.

To illustrate this heat load measurement in section \ref{sec:heating}, we'll start with the same configuration that was used to validate the method and its calibration in section \ref{calibration}. The sample is thermally contacted with silver lacquer and is also cooled by the gas inside the wet cryostat. The experiment will be reproduced in the dry cryostat (sample in vacuum) with different types of joints, silver lacquer for comparison, GE varnish and Apiezon-N grease (see \ref{sec:conductance}). In the dry cryostat, the magnetic field is produced by a permanent magnet and varies a little depending on the exact relative position of the crystal from 203\,mT to 215\,mT for different configurations, slightly higher than the reference value of 184\,mT used in \ref{sec:methods} but still comparable. Despite the compactness of the assembly, this justifies all the more the calibration step described in section \ref{calibration} for each configuration.

\subsection{Thermometry under external laser heating}\label{sec:heating}

\begin{figure}[ht!]
\centering
\includegraphics[width=.8\columnwidth]{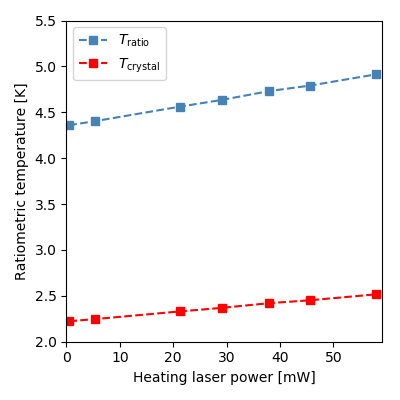} 
\caption{$T_\mathrm{ratio}$ as a function of the heating laser power. From the calibration procedure described in \ref{calibration}, we can infere the crystal temperature $T_\mathrm{crystal}$ for each measurement of $T_\mathrm{ratio}$. }
\label{fig:heating}
\end{figure}

As mentioned, we add to the previous configuration in \ref{calibration} a heating laser that is essentially collinear and counterpropagating with the ratiometric temperature probe beam (See Fig.\ref{fig:mount}). There is no interplay between the heating and the probe beams in the crystal since they are well separated spectrally. Because of the large power of the heating laser, a weak parasite is observed on the probe detector. This may influence the temperature measurement so we subtract it before processing $T_\mathrm{ratio}$. We vary the heating power and observe the dependency in Fig.\ref{fig:heating}.

The ratiometric temperature, as measured, has been calibrated in \ref{calibration}. So from the curve in Fig.\ref{fig:calibration}, we can estimate the crystal $T_\mathrm{crystal}$ for each measurement of $T_\mathrm{ratio}$.

We can clearly see that the temperature of the crystal increases with heating. The dependency is essentially linear with a slope of 5.1\,K per watt of heating laser. This coefficient has the unit of a thermal resistance.

This validates the method under realistic experimental conditions, but for the moment it says nothing about the thermal cooling of the sample. The first step is to estimate the light power actually deposited as heat. As the crystal is very transparent, regardless of the erbium dopants, it is difficult to estimate its absorption by comparing the incoming and outgoing powers, the latter also including reflection losses at the interfaces. However, we were able to measure the absorbed power in the dry cryostat, which has a thermometer precise enough to observe a temperature rise at maximum heating power. This indicates that around 4\% of the incident power is absorbed. So we expect that the crystal thermal resistance obtained from Fig.\ref{fig:heating} is actually 130\,K per watt of absorbed laser. Finally the thermal conductance is usually referenced to the surface in contact with the cold plate, namely 5\,mm$\times$4\,mm for the base of our crystal. 
This definition is debatable in the case of the wet cryostat, as the free surface of the crystal is also cooled by the gas. Since it is not possible to distinguish between the two cooling mechanisms, and in order to be able to compare with the corresponding measurements in the dry cryostat (sample in vacuum), we will keep the base surface of the crystal as a reference (5\,mm$\times$4\,mm) and discuss this point again in the final analysis. Under this assumption, we find a thermal conductance of $3.9\times 10^{-2}$ W\,K$^{-1}$\,cm$^{-2}$ at the crystal base.

Our aim is to make a comparison between different configurations in section \ref{sec:conductance}, but before doing so it is worth discussing this value quantitatively. The heat conduction across solid/solid interfaces with different kind of joints has been discussed comparatively by Ekin \cite[Fig. 2.7]{Ekin}. The range
$10^{-1}-10^{-2}$ W\,K$^{-1}$\,cm$^{-2}$ seems a good order of magnitude for solder of layered joints, without prejudging the nature of the materials in contact and the applied pressure. In that sense, our measurement gives a correct order of magnitude if the conductivity is ensured by the silver lacquer. On the other hand, estimating conduction by gas is much more speculative because of convection around the sample. However, we can give a qualitative answer by reproducing the same measurement with the sample in a vacuum, as we shall now see.

\subsection{Thermal conductance evaluation}\label{sec:conductance}
The aim here is not to push thermal conductance metrology further in terms of precision, but rather to validate an original method under real conditions.

\begin{figure}[ht!]
\centering
\includegraphics[width=\columnwidth]{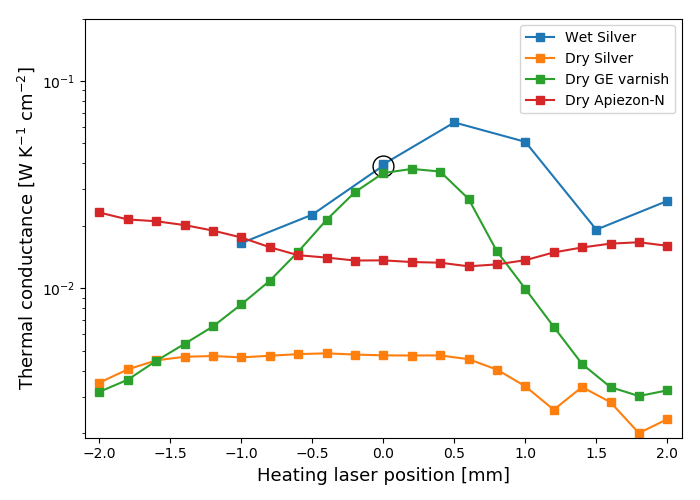} 
\caption{Thermal conductance measured by optical thermometry in four different configurations: wet or dry cryostat with silver lacquer, GE varnish or Apiezon-N grease joints as abbreviated in the legend. We vary the position of the heating laser from the probe laser referenced at 0\,mm. The crystal size along this dimension is 5\,mm. The circled marker corresponds to the reference measurement described in section \ref{sec:heating} (position 0\,mm, Wet Silver curve).}
\label{fig:conductance}
\end{figure}

We are now going to explore three new configurations compared with the previous experiment (silver lacquer thermal contact  in a wet cryostat) that serves as our reference, in the dry cryostat (sample in vacuum) with three types of joints, silver lacquer for comparison, GE varnish and Apiezon-N grease, and extract the thermal conduction. In order to explore a possible thermal gradient and inhomogeneities in the crystal, we additionally move the heating laser position away from the probe beam (defining the position 0\,mm). The results are presented in Fig.\ref{fig:conductance} and they can be analysed as follows, with three points of view.

\subsubsection{Heating spatial dependence}
The aim of measuring the spatial variation in heating is to investigate the existence of a gradient within the crystal. The curves have a chaotic appearance and cannot be considered neither convex nor concave systematically. It is impossible to draw any conclusions. The curvature is probably not linked to a variation in conductance, but more likely to a variation in absorbed power, which we assume to be constant (4\% of the incident power). The absorbed power can vary greatly when the heating beam is translated. Several phenomena are possible.
The beam (diameter about 1\,mm) touches the edge of the crystal (width 5\,mm), encounters an inclusion in the crystal or a bonding residue. It is then difficult to predict whether the absorbed power decreases (convex curves in Fig.\ref{fig:conductance}) as it is clipped by the crystal or increases (concave curves) as it is diffused inwards. This explains the chaotic nature of the curves.
In any case, we do not observe any systematic effect of inhomogeneous heat diffusion in the crystal. This is difficult to imagine anyway, as the typical conductivity of dielectric materials comparable to \yso, such as sapphire or quartz, is around  $10^{3}$ W\,K$^{-1}$m$^{-1}$  (see Fig. 2.1 of Ekin \cite{Ekin}) i.e. 10\,W\,K$^{-1}$ over a typical dimension of one cm. This is at least two orders of magnitude greater than the conductance of the interface(s) ($10^{-1}-10^{-2}$ W\,K$^{-1}$\,cm$^{-2}$).

\subsubsection{Qualitative differences between thermal joints}
In the end, the three bonding techniques gave similar results in terms of conductance. A summary of the different values can be found tabulated in Table~\ref{fig:tab}. Silver lacquer (sample in vacuum) nevertheless appears to perform less well than Apiezon-N grease or GE varnish. We suggest that varnish should be preferred, as its bonding is more robust than grease. With grease, we observed sample detachment during the colddown. Varnish is therefore an excellent compromise between thermal conductivity, ease of use and bond strength.


\begin{table}
\caption{\label{fig:tab} Summary of the typical conductance of the different thermal joints (in W\,K$^{-1}$\,cm$^{-2}$). The average of the conduction is taken from the different positions plotted in Fig.\ref{fig:conductance}. }
\begin{ruledtabular}
\begin{tabular}{cccc}
Cryostat & Sample in & Thermal joint & Average conductance\\\hline\hline
Wet & gas &  silver lacquer  & $3.5\times10^{-2}$ \\\hline
Dry & vacuum & silver lacquer & $4.0\times10^{-3}$ \\\hline
Dry & vacuum & GE varnish & $1.4\times10^{-2}$ \\\hline
Dry & vacuum & Apiezon-N grease & $1.6\times10^{-2}$ \\
\end{tabular}
\end{ruledtabular}
\end{table}

\subsubsection{Solid/solid vs solid/gas conduction}
To conclude, it is possible to compare the curves with silver lacquer with sample in vacuum (orange curve in Fig.\ref{fig:conductance}, average conductance of $4.0\times10^{-3}$\,W\,K$^{-1}$\,cm$^{-2}$) or in gas (blue curve in Fig.\ref{fig:conductance}, average of $3.5\times10^{-2}$\,W\,K$^{-1}$\,cm$^{-2}$ ). If we assume that the quality of the contact with the lacquer is the same in both cases, then the difference in conductance can only be explained by the gas cooling process whose efficiency is an order of magnitude higher.

This is interesting information that can be compared with the literature. The conductivity of the helium gas is well known down to 2\, K \cite{Barron2017, HANDS1981697}. We will take $10^{-4}$ W\,K$^{-1}$\,cm$^{-1}$ as a good order of magnitude at 4K. To compare with the conductance measurements  in Fig.\ref{fig:tab}, we need to determine the distance between the sample holder and the crystal, as the minimal gas conductance distance. To do this, we'll take 3\,mm, the height of the sample. We therefore expect a conductance of $3.3\times10^{-4}\,$W\,K$^{-1}$\,cm$^{-2}$. This is actually two orders of magnitude lower than what we are measuring that is more in the range of $10^{-2}\,$W\,K$^{-1}$\,cm$^{-2}$.

The cooling by the gas is therefore more efficient than the simple conduction mechanism. It undoubtedly relies mainly on convection, which remains difficult to evaluate outside a dedicated test bench. However, our analysis does provide some answers in a practical user case.

\section{Conclusion}\label{ccl}

We have implemented a new optical thermometry technique specifically dedicated to cryogenic temperatures. It is based on spectroscopic measurements of Boltzman populations of the erbium spin levels. Despite its very general nature, this method does not allow an absolute temperature measurement, but requires a calibration step. It retains the advantage of making a local temperature measurement of the sample without additional electrical contacts.

It is this property that we are finally exploiting to make measurements of the interface conductance with the cold cryostat plate. An auxiliary laser is used for direct heating of the sample.

The profusion of rare earth ion levels opens up possibilities in different temperature ranges. As mentioned in the introduction, the crystal field levels that distribute luminescence over several peaks are already being exploited to make room temperature measurements, with applications in biology, for example.
As we have shown, the electron spin levels are well adapted to the cryogenic range around 4K.
It would be interesting to reproduce our study for the mK range, this time using the hyperfine structure. Although such an analysis will inevitably raise fundamental questions of spin thermalisation, it would allow to study a range that is increasingly used with the democratisation of dilution cryostats and where standard thermometry remains very delicate. 

\section*{Acknowledgements}
The authors acknowledge support from the French National Research Agency (ANR) through the project MARS No. ANR-20-CE92-0041, from the Plan France 2030 project QMEMO No. ANR-22-PETQ-0010 and from the ANRT (Association nationale de la recherche et de la technologie) with a CIFRE fellowship No.  2021/0803.


%
%

%


\bibliography{art_Boltzman_thermometry}

\end{document}